\documentclass[11pt,twoside]{article}

\usepackage{asp2006}
\usepackage{epsf}
\usepackage{psfig}
\usepackage{lscape}
\usepackage{mathabx}

\begin{document}

\def\kms{\hbox{km~s$^{-1}$}}
\def\kpc2{\hbox{kpc$^{-2}$}}
\def\msun{\hbox{$\hbox{M}_{\odot}$}}
\def\lsun{\hbox{$\hbox{L}_{\odot}$}}
\def\cm2{\hbox{cm$^{-2}$}}
\def\HI{\hbox{{\rm H}\kern 0.1em{\sc i}}}
\def\HII{\hbox{{\rm H}\kern 0.1em{\sc ii}}}

\title{Tidal Tails in Interacting Galaxies: Formation of Compact Stellar Structures}
\author{B. Mullan\footnotemark[1], J. C. Charlton\footnotemark[1], I. S. Konstantopoulos\footnotemark[1], N. Bastian\footnotemark[2], R. Chandar\footnotemark[3] , P. R. Durrell\footnotemark[4], D. Elmegreen\footnotemark[5], J. English\footnotemark[6], S. C. Gallagher\footnotemark[7], C. Gronwall\footnotemark[1], J. E. Hibbard\footnotemark[8], S. Hunsberger\footnotemark[1], K. E. Johnson\footnotemark[9], A. Kepley\footnotemark[9], K. Knierman\footnotemark[10], B. Koribalski\footnotemark[9], K. H. Lee\footnotemark[1], A. Maybhate\footnotemark[11], C. Palma\footnotemark[1], W. D. Vacca\footnotemark[12]}

\footnotetext[1]{Pennsylvania State University, Department of Astronomy \& Astrophysics, 525 Davey Lab University Park PA 16803; mullan@astro.psu.edu}
\footnotetext[2]{University of Cambridge, Institute of Astronomy, Madingley Road, Cambridge CB3 0HA, UK}
\footnotetext[3]{The University of Toledo, Department of Physics and Astronomy, 2801 West Bancroft Street, Toledo, OH 43606}
\footnotetext[4]{Youngstown State University, Department of Physics and Astronomy, Youngstown, OH 44555}
\footnotetext[5]{Vassar College, Department of Physics \& Astronomy, Box 745, Poughkeepsie, NY 12604}
\footnotetext[6]{University of Manitoba, Department of Physics and Astronomy, Winnipeg, Manitoba R3T 2N2, Canada}
\footnotetext[7]{The University of Western Ontario, Department of Physics and Astronomy, 1151 Richmond Street, London, Ontario, N6A 3K7, Canada }
\footnotetext[8]{National Radio Astronomy Observatory, 520 Edgemont Road, Charlottesville, VA 22903-2475B}
\footnotetext[9]{University of Virginia, University of Virginia, P.O. Box 3813, Charlottesville, VA 22904}
\footnotetext[10]{Arizona State University, School of Earth and Space Exploration, Bateman Physical Sciences Center F-wing Room 686, Tempe, AZ 85287-1404}
\footnotetext[11]{ Space Telescope Science Institute, 3700 San Martin Drive, Baltimore, MD 21218}
\footnotetext[12]{Stratospheric Observatory for Infrared Astronomy/ Universities Space Research Association, NASA Ames Research Center, MS 144-2, Moffett Field, CA 94035}

\begin{abstract}

We have used F606W ($V_{606}$)- and F814W ($I_{814}$)- band images from the \textit{Hubble Space Telescope} (\textit{HST}) to identify compact stellar clusters within the tidal tails of twelve different interacting galaxies. The seventeen tails within our sample span a physical parameter space of HI/stellar masses, tail pressure and density through their diversity of tail lengths, optical brightnesses, mass ratios, HI column densities, stage on the Toomre sequence, and tail kinematics. Our preliminary findings in this study indicate that star cluster demographics of the tidal tail environment are compatible with the current understanding of star cluster formation in quiescent systems, possibly only needing changes in certain parameters or normalization of the Schechter cluster initial mass function (CIMF) to replicate what we observe in color-magnitude diagrams and a Brightest M$_V$ -- log N plot.
\end{abstract}

\section{Introduction}

Galaxy interactions are well known to induce bursts of star formation within the central kiloparsecs of galaxies and within their tidally-distorted features (e.g.\ \citealp{schombert}; \citealp{hibvan}; \citealp{duc}). It has been previously suggested that star formation is manifested in different kinds of stellar ``packaging," like isolated stars, bound groups, loose associations, globular clusters, and dwarf galaxies \citep{K03}. It is now evident, however, that clustered star formation is the sole product of star forming activity (\citealp{lada}; \citealp{fall}), shaped by a number of disruption mechanisms to produce field stars and associations. 

Tidal tails have lower extinctions and less complicated star formation histories -- and resulting stellar populations -- than galaxy interiors. They therefore present a unique opportunity to study a relatively observationally simple, yet physically interesting environment, and assess how it shapes cluster sizes, color and magnitude distributions, and statistical properties. We may also compare what is observed in these environments to prior studies of galaxy interiors. We report our recent work in elucidating this issue with general diagnostics of cluster candidates detected in \textit{HST} WFPC2 images. This represents a pilot study of using a large sample of galaxies with limited bandpass coverage to uncover the physics of star cluster formation and survival; a test of how to do more with less. 

\section{Observations and Reductions}

The sample of tidal tails is presented in Figure 1 as Sloan Digital Sky Survey images with WFPC2 fields of view overlaid. These systems were selected to represent a range of optical brightnesses, merger ages, HI column densities and masses, interaction mass ratios, and presence of tidal dwarf candidates (Mullan et al.\ in preparation).

A full description of observations and data reduction procedures for these systems will be presented in Mullan et al.\ (in preparation). In summary, star cluster candidates were detected using the DAOFIND program in the IRAF package DAOPHOT. The criteria for selection and follow-up extinction- and charge transfer efficiency-corrected photometry (\citealp{schlegel}; \citealp{cteweb}) with the PHOT task of APPHOT were chosen to emulate \citet{K03}, including the criterion $V-I$ $<$ 2.0. This allows populations of old, metal poor globular clusters ($V-I$ $\sim$ 1) akin to those observed in the Galaxy \citep{Reed}, as well as metal-rich clusters with $V-I$ $<$ 1.5 (e.g.\ \citealp{Kundu}). This also allows solar metallicity globular clusters with $V-I$ $\sim$ 1.2--1.3 (Lee, Lee, \& Gibson 2002).

The tail regions were defined in $V_{606}$- band, Gaussian-smoothed images as contours set to one count above the average background, corresponding to 24.3 -- 25.7 mag arcsec$^{-2}$ in $V_{606}$. Central galaxies, where present in the images, were excluded based on where their radial brightness profiles indicated anti-truncated disk debris \citep{erwin}. Our detections are 50\% complete to $V_{606} \approx$ 25.7, and K-S tests indicate that in-tail sources are not likely drawn from the same population as the out-of-tail sources (p $<$ 0.1 in most cases).

\begin{figure}[htbp]
\plotone{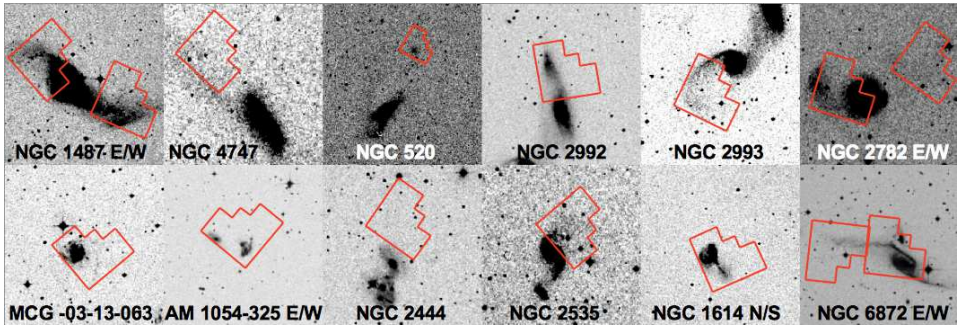}
\caption{\textit{Sloan Digital Sky Survey} images of the Galaxies used in this project, with \textit{HST} fields of view overlaid. The distinctions between eastern and western or northern and southern tails is indicated by separate pointings (NGC 1487 and NGC 2782), or separate regions within a single image (AM 1054-325 and NGC 1614). North and east are up and to the left, respectively. Images are stretched and scaled individually to enhance tidal features.}

\end{figure}

\section{Discussion}

Figure 2 shows an image of a representative system, NGC 2992, with in-tail cluster candidates indicated with circles and squares for out-of-tail objects meeting the same selection criteria. A color-magnitude diagram for the in-tail objects is also presented, with photometric model tracks of 10$^6$, 10$^5$, and 10$^4$ M$_{\odot}$  simple stellar populations with instantaneous-burst star formation rates \citep{BC03} overlaid. Although two photometric bands are insufficient to draw robust conclusions on individual masses and mass functions, rough estimates and limits can be made. In all systems, V$_{606}$ and $V_{606}-I_{814}$ indicate that the most massive clusters are $\sim$ 10$^6$ M$_{\odot}$. We do not find clusters in our sample with masses $\gtrsim$ 10$^7$ M$_{\odot}$.

\begin{figure}[htbp]
\plotone{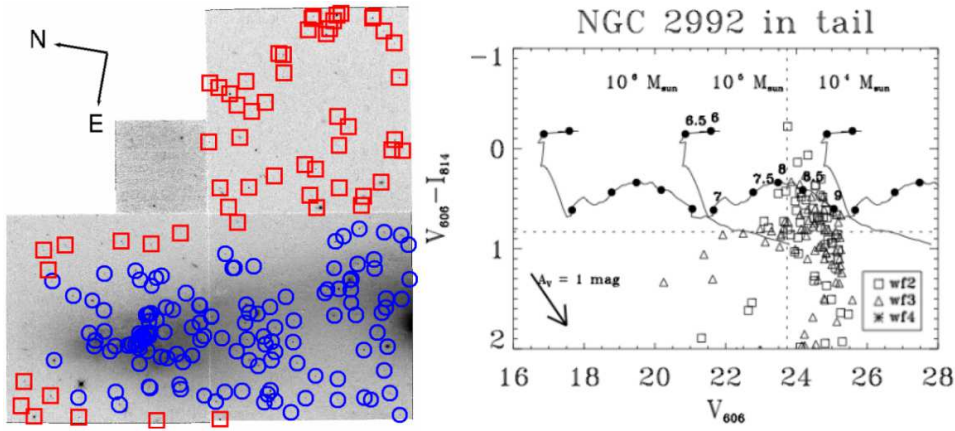}
\caption{Left: WFPC2 mosaic of NGC 2992 debris, with in-tail cluster candidates indicated with circles and out-of-tail sources matching the selection criteria in squares. Right: color-magnitude diagram for in-tail sources. The vertical dotted line indicates where M$_{V}$ = -8.5, and the horizontal line where $V-I$ = 0.7 (both are Johnson-Cousins). Objects to the left and above these respective demarcations are more likely to be clusters. Sources detected in 2nd, 3rd, and 4th chips of the WFPC2 are represented as different symbols as shown on the plot. No corrections are made for incompleteness. \citet{BC03} SSP model tracks are included, with log(age) from 6--9. The extinction vector is labeled. }
\end{figure}

By counting the number of in-tail cluster candidates per unit area of the tail, and subtracting the number of sources found with the same selection criteria outside the tail per unit area of non-tail WFPC2 regions, we find a cluster candidate surface density for each system. Figure 3 compares our results to those of \citet{K03}, given an additional M$_V$ $<$ -8.5 (Johnson-Cousins system) cutoff. On the surface, our results are compatible with \citet{K03} who find similar values, though we contend that accounting for rapid cluster fading with age by extending the limit to fainter magnitudes and improving upon background subtraction reveals clusters in every star-forming tail (Mullan et al., in preparation).

\begin{figure}[htbp]
\plotone{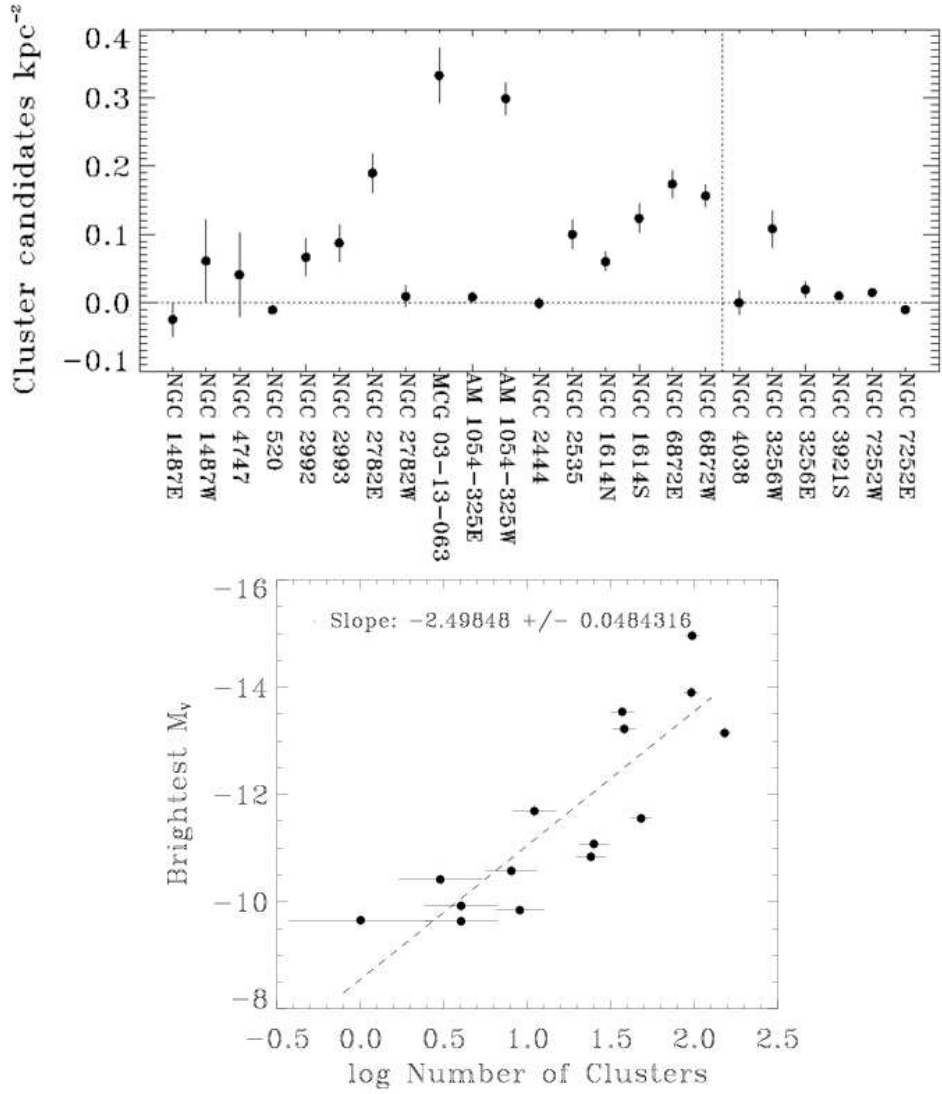}
\caption{Top: cluster candidate (M$_V <$ -8.5, $V-I <$ 0.7) excess for all tails. The points to the left of the vertical dotted line comprise our sample; those to the right are from \citet{K03}. Bottom: Brightest cluster M$_V$ vs. log number of cluster candidates (M$_V <$ -9).  }
\end{figure}

Also provided is a typical statistical diagnostic for the cluster populations, a plot of the brightest M$_{V, 606}$\footnotemark[1] vs. the logarithm of the number of clusters detected for each system. The slope of this correlation, -2.50 $\pm$ -0.05, is similar to the -2.3 $\pm$ 0.2 value found for other extragalactic environments \citep{whitmore07}. \citet{larsen09} suggests that this statistical size-of-sample effect can be produced from a Schechter CIMF with a exponential slope of -2 and cutoff (M$_*$) mass of $\approx$ 2.1 $\times$ 10$^5$ M$_{\odot}$, given cluster disruption mechanisms from gas expulsion, relaxation, stellar evolution, and the external environment (e.g.\ \citealp{whitmore07}; \citealp{bastian08}). In many cases such as NGC 6872E and AM 1054-325W, the profusion of sources between 10$^5$ -- 10$^6$ M$_{\odot}$ may indicate higher star formation efficiencies from high GMC compression by a turbulent ISM \citep{elm97} or through a similar magnetohydrodynamic effect \citep{padoan} in these ``ideal" star-forming tails. This may also affect the luminosity function, whose luminosity-dependent logarithmic slope of -2.5 as sampled in this V$_{606}$ range (see the contribution by Gieles; \citealp{gieles09}) may change with the altered CIMF. But without any clusters of mass 10$^7$ M$_{\odot}$, the star formation efficiencies in these tails may not be as high as in the Antennae, M82, NGC 1316, NGC 7252, or Arp 220 (\citealp{zhang}; \citealp{smith07}; \citealp{bastian06}; \citealp{wilson06}).

\footnotetext[1]{Transformations from M$_{V,606}$ to M$_V$ have not been completed in this case, but are not expected to alter the slope appreciably.}

\section{Conclusions}

Preliminary evidence suggests that star cluster populations are governed by similar luminosity and initial mass functions as their cousins in field galaxies. Our set of 17 tails show a maximum cluster mass of $\sim$10$^6$ M$_{\odot}$, while the Brightest M$_{V,606}$ vs.\ log N plot is consistent with both mass-dependent and mass-independent prescriptions of cluster disruption (``infant mortality"), with subsequent dissolution from 2-body relaxation and stellar evolution for clusters drawn from a Schechter CIMF. Such a CIMF, if it had a higher cutoff mass, would imply higher star formation efficiencies, aided by turbulent pressure in the tidal tails. Future work will address these areas through further analysis of the color, magnitude, and size distributions of these clusters, as well as the kinematics and ISM composition of their surrounding environments.  

\acknowledgements This projected was supported by a grant from the Space Telescope Science Institute (grant no.\ HST-GO-11134.05-A).

\end{document}